\documentstyle[12pt]{article}
\newcommand{\be}{\begin{equation}}
\newcommand{\ee}{\end{equation}}
\newcommand{\bea}{\begin{eqnarray}}
\newcommand{\eea}{\end{eqnarray}}
\newcommand{\ba}{\begin{array}{l}}
\newcommand{\ea}{\end{array}}

\newcommand{\bb}{\begin{thebibliography}}
\newcommand{\eb}{\end{thebibliography}}
\newcommand{\ci}[1]{\cite{#1}}
\newcommand{\lab}[1]{\label{#1}}
\newcommand{\re}[1]{(\ref{#1})}

\newcommand{\half}{{\textstyle{\frac{1}{2}}}}
\newcommand{\cM}{{\cal M}}

\begin{document}
\begin{center} \large \bf {A test of the instanton vacuum chiral
quark model
with axial anomaly low-energy theorems }
\end{center}
\begin{center} \large
 M.M. Musakhanov$^{1}$   and
F. C. Khanna $^{2}$
\end{center}
\begin{center}
\begin{description}
\item [\tt $1$]
Theoretical  Physics Dept, Tashkent State University, 700095,
 Uzbekistan
\item [\tt $2$]
  Department of  Physics, University of Alberta, Edmonton, Canada
 $T6G 2J1$ \\
and TRIUMF, 4004 Wesbrook Mall, Vancouver, BC,
 Canada, $V6T 2A3 $ \\
 E-mail: yousuf@iaph.silk.glas.apc.org, khanna@phys.ualberta.ca
\end{description}
\end{center}
\centerline {\large Abstract }
The QCD$+$QED axial anomaly low-energy theorems are applied to
estimate the
accuracy of
an instanton vacuum - based chiral quark model. The low-energy
theorems
give an  exact relation between  the
 matrix elements
of the gluon and photon parts of the QCD$+$QED
axial anomaly operator equation.
The matrix elements between
vacuum and two photon states and between vacuum
and two gluon
states are calculated  for  arbitrary $N_f .$
It is shown that this model does satisfy the low-energy theorems
with an accuracy of $ \sim  17\%$.

We estimate also the contribution of the
nonperturbative  conversion of gluons into photons to the
decay $\eta ' \rightarrow 2 \gamma $ and compare with experimental
data.\\
\vskip 5mm
PACS number(s):  11.15.Tk ,  12.38.Lg. \\
\newpage
\section { Introduction}

In gauge theories the symmetries of the initial classical lagrangian
may be destroyed by quantum fluctuations. One of the most important
examples
is the famous axial anomaly.
The axial anomaly leads to many interesting nonperturbative phenomena
in physics. Among them are $B$-violation processes in electroweak
$(EW)$ physics, $U_A (1)$ problem in QCD etc.
The solution of these problems  is intimately related to the
topologically nontrivial structure of the vacuum in the gauge
theories.

In this paper we apply the axial anomaly low-energy theorems to test
the chiral quark model which is based on the instanton model of QCD
vacuum.
We find that this model does satisfy these theorems with an  accuracy
of $ \sim 17 \%$.
This conclusion provides solid background to calculate  different
amplitudes of nonperturbative conversion of  gluons  into hadrons and
photons.

We estimate also the contribution of the
nonperturbative  conversion of gluons into photons to the
decay  $\eta ' \rightarrow 2 \gamma $ .
 By using experimental date for the width of
$\eta ' \rightarrow 2 \gamma$  we
conclude that $\eta '$-singlet axial current coupling constant
is given as $f_{0} =1.16 f_{\pi}.$

\subsection{ Low-energy theorems from axial anomaly}

The axial anomaly in the divergence of the singlet axial current in
QCD $+$ QED leads to a  low-energy  theorem for  the matrix elements
of this operator equation over vacuum and two--photon states:
\be
 \langle 0|N_f \frac{g^2}{32\pi^2}G\tilde G | 2\gamma \rangle =
N_c \frac{e^2}{8\pi^2} \sum_{f}{Q^{2}_{f}} F^{(1)}\tilde F^{(2)}
\lab{theorem}
\ee
at $q^2 = 0$ \ci{Shi}.
Here,  $N_f$ is the number of the flavors,  $g$ is the QCD coupling
constant
with
$G\tilde G= \frac{1}{2}\epsilon^{\mu\nu\lambda\sigma}
G^{a}_{\mu\nu}  G^{a}_{\lambda\sigma}$ and $G^{a}_{\mu\nu}$ being the
operator of the gluon field strengths , $N_c$ is the number of the
colors,
 $e$ and $Q_{f}$ are QED coupling constant and the electric charges
of
the quarks respectively,
$2F^{(1)}\tilde F^{(2)} = \epsilon^{\mu\nu\lambda\sigma}
 F^{(1)}_{\mu\nu}  F^{(2)}_{\lambda\sigma},$
$F^{(i)}_{\mu\nu} = \epsilon^{(i)}_\mu q_{i\nu} -
\epsilon^{(i)}_\nu q_{i\mu},$  $\epsilon^{(1,2)}_\mu$,
$q_{1,2}$ are polarizations and momenta of  photons
respectively and $q=q_{1} + q_{2}$. This
relation is a consequence of the absence of a massless singlet
pseudoscalar boson.
Here the contribution of the quark masses is neglected.

As it is evident, gluons can interact with photons only through quark
loops.
In perturbation theory it  leads to at least $\sim g^{4}$ results for
the
left hand side of  Eq. \re{theorem}
(see  e.g. recent discussion of the higher-loop contributions to the
axial anomaly \ci{FMP}).  So, the solution of this theorem is related
only to  the nonperturbative phenomena connected with the
structure of QCD vacuum.

Another nontrivial low-energy theorem concerns the matrix element
over
vacuum and two-gluon states:
\be
 \langle 0| g^2 G\tilde G | 2 gluons \rangle = 0
\lab{theorem1}
\ee
at the  limit $q^2 = 0.$

These matrix elements are calculated in the instanton vacuum
generated
N-JL type quark model \ci{DP86}, \ci{DPW95} for  arbitrary $N_f .$

\subsection { The instanton vacuum of QCD}

The instanton is the solution of  gluodynamics in the
Euclidian space\ci{BPST75}:

\be
    A_{\mu}^{Ia}(x) = 2g^{-1}O_{I}^{ab}\bar\eta_{\mu\nu}^{b}
\frac{\rho^2 (x-z)_\nu }{\left[(x-z)^2 + \rho^2 \right] (x-z)^2 }.
\lab{instantonx}
\ee
The anti-instanton solution has the same form as in Eq.
\re{instantonx} but with
 $\eta_{\mu\nu}^{b}$
 instead of $\bar\eta_{\mu\nu}^{b}$.
Here $O_{I(\bar I)} $ is the   orientation  matrix of the instanton
 $I$ (anti-instanton $\bar I$) in color space,
$\bar\eta (\eta )$ -- t'Hooft factors \ci{tH76},
$\rho$ is the size and $z$ is the position of the instanton.
 For large interinstanton distances $R>>\rho$ the sum of  the $N_+$
instantons
and the $N_-$ anti-instantons is also an approximate solution.
The calculation of the
action for gluon fields leads to a sum of  actions of  free
instantons and classical interinstanton potential
$V(R, \rho_1 , \rho_2 , O )$, where
$O=O^{T}_{1}O_{2} $ is a matrix of relative orientation.
The most important part is  the
instanton-antiinstanton potential  $V_{I\bar I}$.
  It is well--known that  at large distances it has a form of
the dipole-dipole interaction potential
and may be attractive.

The main assumption of the model concerns small distances $R \sim
\rho$.
At small distances it is assumed
 that there is a repulsion.
This assumption  is  supported
by both   phenomenological
 and theoretical considerations
\ci{Shu82}, \ci{DP84}.
This  leads to a stabilization of the size and density of instantons.
The quantities  mean number of
instantons $\langle N\rangle$ and a mean instanton size
$ \overline{\rho}$ include
 effects of the instanton interactions  \ci{DPW95}.
   In the
  following calculations, for simplicity,
all instanton sizes are considered to be  $\bar\rho$.

Both the phenomenological estimates  and variational calculations
lead to a mean
interinstanton distance  of
$\bar R =  \left(\frac{V}{{\langle N\rangle}}\right)^{1/4} \sim 1\,
fm$
and
a mean instanton size of $\bar\rho \sim 1/3 \, fm.$ The
small packing parameter $(\bar\rho / \bar R)^{4} = 0.012$
provides a possibility for the independent averaging over positions
and orientations
of instantons.

\subsection { The chiral quark model }

The main assumption of this model is an  interpolation formula for
the
quark propagator in the single instanton field.
 It is approximated as the sum of a free propagator
and an explicit contribution of the zero mode \cite{DP86},
\be
\left(i\hat\nabla (\xi_{I(\bar I)} ) + im \right)^{-1}_{\rm 1-inst}
\approx
(i\hat\partial )^{-1} \;\; - \;\; \frac{\Phi_\pm (x; \xi_{I(\bar I)}
)
\Phi_\pm^\dagger (y; \xi_{I(\bar I)} )}{im} .
\label{prop_zero_mode}
\ee
Here, $\hat\nabla = = \hat\partial  - i g \hat A$,
$\Phi_\pm (x ; \xi_{I(\bar I)})$ is the zero mode wave function of
the fermion
 in the background of one $I (\bar I )$\ci{tH76}. It  depends on the
collective instanton variables $ \xi_{I(\bar I)}$ -- the size $\rho$,
the position $z$ and the orientation $O$ of the instanton.

This interpolating formula should be accurate both at small momenta
($p \ll 1/\bar\rho$), where the zero mode is dominant, and
at large momenta ($p \gg 1/\bar\rho$), where the propagator reduces
to the
free one. In the background of an
$N_\pm$--instanton configuration  and keeping in mind the low density
of
the instanton media  this formula leads to the partition function of
the model:
\be
 Z_N = \int D\psi D\psi^\dagger \exp  (\int d^4 x \, \psi^\dagger i
\hat\partial  \psi )  \,  W_{+}^{N_+}  \, W_{-}^{N_-},
\lab{Z_NW}
\ee
where
\be
W_\pm = \left( - \frac{4\pi^2\bar\rho^2}{N_c} \right)^{N_f}
\int \frac{d^4 z}{V} \det J_\pm (z),
\ee
\be
J_\pm (z)_{fg} = \int \frac {d^4 kd
^4 l}{(2\pi )^8 } \exp ( -i(k - l)z)
\, F(k) F(l) \, \psi^\dagger_f (k) \half (1 \pm \gamma_5 ) \psi_g (l)
,
\lab{J_pm}
\ee
and the contribution of the current quark masses is neglected.
The form-factor $F$ is related  to the zero--mode wave function
 in momentum space $\Phi_\pm (k; \xi_{I(\bar I)}) $ and is equal to:
\be
F(k) = - t \frac{d}{dt} \left[ I_0 (t) K_0 (t) - I_1 (t) K_1 (t)
\right] \; \rightarrow
\left\{ \ba 1 \, \, \, \,
t \rightarrow 0 \\  \frac{3}{4} t^{-3} \, \, \, \,
 t \rightarrow \infty \ea \right.
\\
, \nonumber
\lab{F(k)}
\ee
with  $t =\frac{1}{2} k \bar\rho.$

The   formula:
\be
 (ab)^N = \int d\lambda \exp (N ln \frac {aN} {\lambda} - N +\lambda
b )
\lab{ab^N}
\ee
 (here $N >>1)$ provides the final expression for the partition
function \ci{DPW95}:

\be
 Z_N = \int D\psi D\psi^\dagger \exp
(\int  \psi^\dagger i \hat \partial  \psi
+ Y_{+} + Y_{-}),
\lab{Z_NY}
\ee
where
\be
Y_{\pm}=  (i )^{N_f} \lambda
\int d^4 z \, \det J_\pm (z) =  \left(\frac{2V}{N}\right)^{N_f - 1}
(i M)^{N_f}
\int d^4 z \, \det J_\pm (z) .
\lab{Y_pm}
\ee
 The  self-consistency condition at the saddle point in Eqs.
\re{Z_NY},
 and \re{Y_pm} leads to
\be
4 N_c V \int \frac{d^4 k}{(2\pi )^4} \frac{M^2 F^4 (k)}{M^2 F^4 (k) +
k^2}
=  N .
\lab{selfconsist}
\ee

\section {  Calculations with the low-energy theorems }

In the quasiclassical(saddle-point) approximation any gluon operator
receives its main contribution from instanton background.
As an example, for one  instanton (anti-instanton)$I (\bar I)$,

\be
{g^2}G^2 (x) =
\frac{192 \rho^4}{\left[ \rho^2 + (x - z)^2 \right]^4} = f(x-z),
\lab{GG}
\ee
and
\be
{g^2}G\tilde G(x) \;\; = \;\; \pm f(x-z) .
\lab{GtildeG}
\ee

In this particular case, the calculation of the matrix element
 can be reduced to the calculation of the partition
 function
\be
\hat Z_{N}  [\kappa , a] =
Z_{N}^{-1} \int  D\psi D\psi^\dagger \exp  (- \hat S_{eff}  ) ,
\lab{hatZ_N}
\ee
 with an
effective action, $\hat S_{eff},$ in the presence of an external
electromagnetic  field, $a_\mu ,$ and an external field
$\kappa (x)$ is given as
\be
\ba
 -\hat S_{eff}  =  \int  \psi^\dagger i \hat D  \psi
+ Y_{+} + Y_{-} + \int dx \left(Y_{G\tilde G +} (x) + Y_{G\tilde G -}
(x)
\right) \kappa (x),\\
Y_{G\tilde G\pm}(x) = \pm \left(\frac{2V}{N}\right)^{N_f - 1} (i
M)^{N_f}
\int d^4 z\, f(x-z) \, \det J_\pm (z) ,
\lab{Y_GtildeGQ}
\ea
\lab{hatS_ef1}
\ee
where $\hat D = \hat\partial  - i e Q_{f} \hat a.$
Finally,  Eq. \re{hatS_ef1} can be rewritten as:
\be \ba
 -\hat S_{eff}  =
\int \psi^\dagger i\hat D \psi +
\\ \\
\left(\frac{2V}{N}\right)^{N_f - 1}
\int dz \,  \det (i  M_{+} (z) J_{+} (z)) +
\left(\frac{2V}{N}\right)^{N_f - 1}
\int dz \, \det (i  M_{-} (z) J_{-} (z)) ,
\lab{hatS_ef3}
\ea \ee
where
$$ M_{\pm} (z) =
 \left( 1 \pm \int dx \kappa (x) f(x-z)\right)^{(N_f - 1)^{-1} }M.$$

Another remarkable formula
\be
\exp (\lambda \det [i A] ) =
\int d\cM \exp\left[ - (N_f - 1) \lambda^{-\frac{1}{N_f - 1}}
(\det\cM )^{\frac{1}{N_f - 1}} + i tr (\cM A) \right]
\lab{expA}
\ee
is used in the following discussion.
It is possible to check this  by  the saddle point approximation  of
the integral.

 By using  Eq.\re{expA} it is easy to show that
the effective bosonized action
$\hat S_{eff}[ \cM_{\pm}, a, \kappa ]$, describing mesons
(the matrices $\cM_{\pm}$)
in the presence of the external fields $a_\mu$ and $\kappa$     is
\be\ba
 -\hat S_{eff}[ \cM_{\pm}, a, \kappa ] = \int dz \left( - (N_f - 1)
\left(\frac{2V}{N}\right)^{ - 1}
(\det\cM_{\pm} )^{\frac{1}{N_f - 1}}\right) + \\ \\
Tr \ln \left( i\hat D + i \cM_{+}MF^2 \left( 1 +
( \kappa  f)\right)^{N_{f}^{-1}} \frac {1}{2}(1+\gamma_5 )  +
 i \cM_{-}MF^2
\left( 1 -  (\kappa f)\right)^{N_{f}^{-1}} \frac {1}{2}(1- \gamma_5 )
\right)
\lab{hat S_ef4}
\ea\ee
For the processes without mesons($\cM_{\pm}=1$) the partition
function is:
\be\ba
\hat Z_{N} [\kappa , a]  =
\exp Tr \
ln ( i\hat D + iMF^2 ( 1 + ( \kappa  f))^{N_{f}^{-1}} \frac
{1}{2}(1+\gamma_5 )  + \\
 iMF^2
( 1 -  (\kappa f))^{N_{f}^{-1}} \frac {1}{2}(1- \gamma_5 ) )
\times ( i \hat \partial  +  iMF^2 )^{-1},
\lab{hatZ_N2}
\ea\ee
where $( \kappa  f) =  \int dx \kappa (x) f(x-z) .$

\subsection { The low-energy theorems for the matrix element between
vacuum
and two-photons states  }

The matrix element  is generated by
$$\frac {\delta \hat Z_{N}  [\kappa , a] }
{\delta \kappa (x) \delta a_\mu (x_1) \delta a_\nu (x_2) }
|_{ \kappa , a = 0}.
 $$
As it is clear from Eq. \re{hatZ_N2} the factors in the vertices in
the corresponding Feynman loop-diagram are $eQ_{f}\gamma_{\mu}$ and
$i M f F^{2}\gamma_{5}N_{f}^{-1}$.

We must calculate ${\Delta (q^2 )}$ (in the
$ \lim {q^2\rightarrow 0}$), which is defined by:
\be \ba
(2\pi )^4 \delta (q - q_1 - q_2 ) \Delta (q^2 ) =
\int dx \exp (-iqx)
 \langle 0|{g^2}G\tilde G(x)| 2\gamma \rangle = \\ \\
e^{2} \epsilon^{(1)}_\mu  \epsilon^{(2)}_\nu \int \langle 0|
T({g^2}G\tilde G(x) j^{em}_\mu (x_1 )
j^{em}_\nu (x_2 ) |0 \rangle \exp i(-iqx + q_1 x_1 + q_2 x_2 ) dx
dx_1 dx_2.
\lab{Delta1}
\ea \ee
 It is clear that
$$\Delta (q^2 ) = \epsilon^{(1)}_\mu  \epsilon^{(2)}_\nu   f(q^2 )
N_c  e^2 \sum_{f}{Q^{2}_{f}} $$
 $$\times Tr [ \int {{d^4 p \over \left( 2\pi  \right)^4}
\frac {iM F_{1}F_{2} \gamma_5
( \hat p - \hat q_1 + iM F^{2}_{1} )
\gamma_{\mu}
(\hat p + iM F^{2}_{3} )
\gamma_{\nu}
(\hat p + \hat q_2 + iM F^{2}_{2} ) }
{ (  (p - q_1 )^2 +  M^2  F_{1}^{4})
(  p^2 +  M^2  F^{4}_{3} )
(  (p + q_{2} )^{2}
+   M^{2}  F_{2}^{4} ) }} $$
\be
+ ( \mu \rightleftharpoons \nu , \, q_1 \rightleftharpoons q_2 )].
\lab{Delta2}
\ee
Here $F_{1}=F((p-q_1 )$, $F_{2}=F(p+q_2 )$, $F_{3}=F(p)$ and
\be
f(q^2 )=\int dx \exp(-iqx) f(x)
\lab{fq^2}
\ee
 is the form-factor of the one-instanton
contribution to ${g^2}G\tilde G$. At $q^2 = 0$
$$f(q^2 =0)= 32\pi^2 .$$
It is easy to show that the trace in Eq.\re{Delta2} can be reduced to
$$
8M^2 \epsilon^{\mu\nu\lambda\sigma}q_{1\lambda}q_{2\sigma}
\Gamma (q^2 , r ),
$$
where
\be
\Gamma (q^2 , r ) =
  \int {{d^4 p \over {\left( 2\pi^4 \right)}}
{  F_{1}F_{2} F_{3}^{2}\over {\left(  (p - q_1 )^2 +  M^2  F^{4}_{1}
\right)
\left(  p^2 +  M^2  F^{4}_{3} \right)
\left(  (p + q_2 )^2 +  M^2  F^{4}_{2} \right)}}} + ...
\lab{Gamma}
\ee
and $r=M\rho$.
It is possible to calculate  this integral analytically
if we put $r=0$ in
Eq.\re{Gamma}(In this case  $F=1).$
In this approximation
and at a small values of $q^{2}, $  we have
\be
\Gamma (q^2 , r=0) =  {1 \over 32\pi^2 M^2 }
\left( 1 - {q^2 \over 12M^2}\right)
\lab{Gamma2}
\ee
As a result,  in the $r=0$ approximation
the left side of the low-energy theorem, Eq.\re{theorem},  is
\be
(N_f \frac{g^2}{32\pi^2})(\frac {4e^2 N_c}{g^2 N_f}
\sum_{f}{Q^{2}_{f}} )
F^{(1)}\tilde F^{(2)}
\lab{theorem2}
\ee
and this coincides with the right side of  Eq.\re{theorem}.

 The calculations of the integrals in  Eq. \re{Gamma}
can be performed also with an  account of the form--factor $F(p)$.
 The deviation of  results of the calculation
from the approximate  results of the $r = 0$ tell us about the
accuracy
of this model.
We need to calculate
\be
\Gamma (q^2 =0 , r) =  M^2
         \int {{d^4 p \over {\left( 2\pi^4 \right)}}
{ F^{2}(p) (F^{2}(p) - p^2 \frac{dF^2 (p)}{dp^2}) \over
{\left(  p^2 +  M^2  F^{4}(p) \right)^3 }}}
\lab{Gamma3}
\ee
 An evolution of  $\Gamma (q^2 =0 , r)$
was performed numerically.
Table 1 presents
the factor $\delta (r) = 1 - \Gamma (0, r)/ \Gamma(0,r=0)$, where
$r=M\rho$.
\begin{table}
{\bf Table 1.} The accuracy of the model        \cite{DP86}
 for the different values of the parameter
$r=M\rho$ in the Eq.   \re{Gamma3}   to satisfy axial anomaly
low-energy theorems.\\
\vskip  2mm

\begin{center}

\begin{tabular}{|c|ccccccccccc|}   \hline
$r$&
 .0  &
 .1  &
 .2  &
 .3  &
 .4  &
 .5  &
 .6  &
 .7  &
 .8  &
 .9  &
1.0  \\   \hline
$\delta(r)$&
.00&
.05&
.09&
.13&
.15&
.17&
.19&
.20&
.21&
.22&
.23\\
\hline
\end{tabular}
\end{center}
\end{table}
\subsection{ The low-energy theorem for the matrix element between
vacuum and two-gluons states }

Here we present, without details,  calculations related to Eq.
\re{theorem1}.
This matrix element can be written in the form:
\be\ba
\langle 0| g^2 G\tilde G | g(\epsilon^{(1)} , q_1 ), g(\epsilon^{(2)}
, q_2 )
\rangle  =
 \epsilon^{(1)a_1}_{\mu_1}  \epsilon^{(2)a_2}_{\mu_2}  \\
\lab{me2} \\
\times\int  \partial^{2}_{2} \,\partial^{2}_{1}\, \langle 0|
T{g^2}G\tilde
G A_{\mu_1}^{a_1}(x_1 ) A_{\mu_2}^{a_2}(x_2)
|0 \rangle  \exp i(q_1 x_1 + q_2 x_2 ) dx_1 dx_2.
\ea\ee
Here $A_{\mu}^{a}(x )$ is a total gluon field,
$\epsilon^{(i)a_i}_{\mu_i} , q_i $ are
the polarization and
 the momentum of gluons respectively.

As usual, we expand the total field $A_{\mu}^{a}(x )$ around  the
instanton background. The main term in Eq.\re{me2} is the
contribution
of the instanton background and is  $\sim O(g^{-2}).$
The next term is the
contribution of the perturbative fluctuations over instanton
 background and is   $\sim O(g^{2}).$
It is easy to see from previous considerations that  $ O(g^{-2})$
term is given by the formula

\be
Z_{N}^{-1} \int D\psi D\psi^\dagger \exp  (- S_{eff} )
\left( \left( Y_{G\tilde GAA +} (x) + Y_{G\tilde GAA -} (x)\right)
Q\right) ,
\lab{GtildeGAAQ2}
\ee
where
\be\ba
Y_{G\tilde GAA \pm} = \pm \left(\frac{2V}{N}\right)^{N_f - 1} (i
M)^{N_f}
\int d^4 z\, f(x-z) \\
\lab{Y_GtildeGAAQ} \\
  \times \int dO
(-\partial^{2}_{1})A_{\mu_1}^{I(\bar I)a_1}(x_1)
(-\partial^{2}_{i})A_{\mu_2}^{I(\bar I)a_2}(x_2)
 \det J_\pm (z) ,
\ea\ee
Here the instanton(anti-instanton) is located at the point $z$ with
its
orientation $O$.

 Repeating  the bosonization trick   leads to the result for the
$ O(g^{-2})$ contribution  which is proportional to
$$ Tr[ (i\hat\partial + iMF^2)^{-1}iM F^2 \gamma_5 ].$$
It is clear that this $Tr$  and as a consequence $O(g^{-2})$ term
are equal to zero.

 The next order   $( O(g^{2})) $ term is the contribution of the
two diagrams. The first diagram is the direct contribution of
the operator
$g^2 G\tilde G$
  which  is equal to
$$- g^2 G^{(1)} \tilde G^{(2)},$$
where $2G^{(1)} \tilde G^{(2)} = \epsilon^{\mu\nu\lambda\sigma}
 G^{(1)a}_{\mu\nu} G^{(2)a}_{\lambda\sigma},$  $G^{(i)a}_{\mu\nu}=
\epsilon^{(i)}_{\mu}  q_{i\nu} - \epsilon^{(i)}_{\nu}
  q_{i\mu}.$

The factors in the vertices of the second loop--diagram are
$g \lambda_{a} /2 \gamma_{\mu}$ and
$i M f F^{2}\gamma_{5}N_{f}^{-1}$. An account of the contribution
from all flavors gives the coefficient $N_{f}.$

A comparison with the  previous calculations  (Eq.\re{Delta1},
\re{Delta2}) leads to the result that the contribution of the second
loop--diagram is
 equal in magnitude but opposite in sign to the contribution of the
first diagram at $q^2 = 0.$

Then,  terms of $ O(g^{2})$ are equal to zero in the  limit
$q^2 \rightarrow 0$.
This conclusion is valid in $r=0$ approximation.
Thus we conclude that the instanton vacuum generated chiral quark
model  satisfies the low-energy theorems with
 an   accuracy given in  Table 1.

\section{ Nonperurbative conversion of gluons into photons and
   $\eta ' \rightarrow 2 \gamma$ decay }
One of the first applications of QED axial anomaly was
electromagnetic decays of pseudoscalar mesons \ci{GJS69}.
Here we consider the application of the QCD$+$QED
axial anomaly to the same kind of decays.
 Let us  consider the matrix element of
the divergence of the axial singlet current
$\partial_{\mu}j^{5}_\mu$ between vacuum and two--photons
 states at nonzero but small $q^2$.
 We will compare this one
 with similar matrix element of the
 divergence of the third component of the axial isovector current
 $\partial_{\mu}j^{5,3}_\mu$.
The definition of currents are
$j^{5}_\mu =\bar q\gamma_\mu\gamma_5 q $ and
$j^{5,i}_\mu = \bar q\gamma_\mu\gamma_5 (\tau_{i}/2) q .$
The matrix elements of these currents
$<0|j^{5}_\mu |\eta '(p)>=3^{1/2}if_0 p_\mu$ and
$<0|j^{5,i}_\mu |\pi_j (p)>= \delta_{ij}if_\pi p_\mu$
defines the couplings $f_0$ and $f_\pi .$

The decay constant of charged pions is $f_\pi = 93.3 MeV .$
In contrast to this one, $f_0$ has not been  measured experimentally.
Our approach provides a  method for  estimating
 $f_0 $ from the data on $\eta ' \rightarrow 2 \gamma$ decay.

The matrix element under consideration is
\be
 \langle 0|\partial_{\mu}j^{5}_\mu | 2\gamma \rangle =
 \langle 0|N_f \frac{g^2}{32\pi^2}G\tilde G | 2\gamma \rangle -
N_c \frac{e^2}{8\pi^2} \sum_{f}{Q^{2}_{f}} F^{(1)}\tilde F^{(2)}
\lab{div}
\ee
We neglect here  the masses of the $u,d,s$ quarks.

The lowest intermediate meson state in this equation is $\eta ' $ -
meson
with the non-zero mass $m_{\eta '}$
due to axial anomaly, as was mentioned above.
With definition
$
 \langle\eta '| 2\gamma \rangle =
f_{\eta '\rightarrow 2\gamma }F^{(1)}\tilde F^{(2)}
$
it is easy to conclude that $O(q^2 )$ terms provide the equation
\be
3^{1/2}f_{0}  f_{\eta '\rightarrow 2\gamma } =
N_c \frac{e^2}{8\pi^2} \sum_{f}{Q^{2}_{f}}
m_{\eta '}^{2} \frac{d(f(q^2 )\Gamma (q^2 , r))}{dq^2}|_{q^2 =0}
\lab{eta'}
\ee
In the $r=0$ approximation we conclude that right side of this
equation
is equal to
\be
N_c \frac{e^2}{8\pi^2} \sum_{f}{Q^{2}_{f}}
(\frac {m_{\eta '}^{2}}{12M^2} + \frac {1}{4}m_{\eta '}^{2}\rho^2 ) .
\lab{rightside}
\ee
On  the other hand, as it is well known,
the QED axial anomaly in the divergence of the
third component of the isovector axial current gives  the expression
\be
f_{\pi} f_{\pi_0 \rightarrow 2\gamma } =
N_c \frac{e^2}{8\pi^2} (Q^{2}_{u} - Q^{2}_{d}),
\lab{pi}
\ee
where $f_{\pi_0 \rightarrow 2\gamma} $ is defined analogously with
$f_{\eta ' \rightarrow 2\gamma }.$
Theoretical value of the width
$\Gamma_{\pi_0 \rightarrow 2\gamma } = 7.25 eV$ is rather close to
the experimental one $7.95 eV.$

With experimental value for the width
$\Gamma_{\eta '\rightarrow 2\gamma } = 4.296 KeV$,
$M=340 MeV$ and $\rho=0.3 fm$ in \re{rightside}
we estimate that
\be
f_{0}  = 1.16 f_{\pi}.
\lab{f_0/f_pi1}
\ee
This quantity was estimated previously by  another method
using $SU_{f}(3)$-symmetry type
arguments and assumptions about the
$J/\psi \rightarrow \gamma \eta (\eta ')$ decay mechanism
\ci{Shu93,Novik80,Ball95}.
Here the axial anomaly equation,  eq.  \re{div}, is used
again but for the matrix elements over the vacuum and $\eta$ and
$\eta '$
states. As a  result, neglecting  the contributions
of the current masses of the light $(u,d,s)$ quarks we get
\be
m^{2}_{\eta '} 3^{1/2}f_{0}=N_{f} \frac{g^{2}}{16\pi^{2}}
<0| G\tilde G|\eta '>
\lab{fetaprime}
\ee
The matrix element on  the right side  can be extracted from
data on the transitions and decays of heavy quarkonium
systems\ci{Novik80}.
Taking into account theoretical relation \ci{Shu93,Novik80} and
experimental widths of the
 $J / \psi \rightarrow \gamma \eta (\eta ') $ decays\ci{Her90}
$$
\frac{|<0| G\tilde G|\eta '>|^2 }
{|<0| G\tilde G|\eta >|^2} \frac{p_{\eta '}^3}{p_{\eta }^3} =
\frac{\Gamma (J / \psi \rightarrow \gamma \eta '(p_{\eta '}))}
{\Gamma (J / \psi \rightarrow \gamma \eta(p_{\eta }) )} = 5
$$
it is easy to find  that \ci{Shu93}
$$
\frac{<0| G\tilde G|\eta '>}
{<0| G\tilde G|\eta >} =2.46 .
$$
On the other hand
$$
N_{f} \frac{g^{2}}{16\pi^{2}}<0| G\tilde G|\eta > =
-2im_{s}<0|\bar s\gamma_{5}s|\eta>=
(\frac{2}{3})^{1/2} f_{8} m^{2}_{\eta} .
$$
$SU_{f}(3)$ symmetry arguments give
$
f_8 = 1.86 f_\pi .
$

This type of the estimates  gives
\be
f_0 = 1.06 f_\pi .
\lab{f_0/f_pi2}
\ee
The estimates, Eqs.\re{f_0/f_pi1} and \re{f_0/f_pi2} agree
with each other even better than the
usual accuracy of the $SU_{f}(3)$-symmetry. It is crucial to repeat
all of these estimates  with account of the $SU_{f}(3)$-symmetry
breaking.

\section{ Conclusion }
The solution of the axial anomaly low-energy theorems
very nontrivially related with nonperturbative instanton
structure of the QCD vacuum.
The  accuracy of the instanton vacuum
based chiral quark model \cite{DP86}
to satisfy these theorems is $ \sim  17\%$ and, probably,
is related with the accuracy of the interpolating formula
\re{prop_zero_mode}
for the quark propagator.

 This approach provides a solid background to calculating the
 different amplitudes of the nonperturbative conversion of  gluons
into hadrons and photons.
As example, we applied  this approach to the
  $\eta ' \rightarrow 2 \gamma$ decay process
and, in result, successfully estimated
$\eta '$-singlet axial current coupling constant,
the Eq.\re{f_0/f_pi1}.

We are planning to apply this approach to
the hadronic and photonic  transitions in a heavy quark systems
and the scattering of these systems on hadrons.

We hope  to provide  further discussions of
 all of these problems in a separate publication.

\section{ Acknowledgments }

One of the authors(M.M.) is grateful to I. Musatov and E. Shuryak
for  useful discussions about  the instantons and acknowledges the
hospitality of the Theory Group of CEBAF, the Theoretical
Physics Institute of the University
of Alberta and the Department of Physics of SUNY at Stony Brook.
The  work of M. Musakhanov is supported in part  by the grant
INTAS-93-0239. The work of F.Khanna is
supported in part by the Natural Sciences and Engineering research
Council of Canada.

\end{document}